\documentclass[aps,prl,twocolumn,floats,floatfix]{revtex4}

\usepackage{amsmath}
\usepackage{graphicx}
\usepackage{natbib}
\usepackage{latexsym}

\usepackage{ifthen}
\usepackage{ifpdf}
\ifpdf
\usepackage{graphicx}
\usepackage{epstopdf}
\else
\usepackage{graphicx}
\usepackage{epsfig}
\fi

\newcommand{\ei}{\hat{a}}
\newcommand{\eidag}{\hat{a}^{\dag}}
\newcommand{\hn}{\hat{n}}

\newcommand{\Lx}{\hat{L}_x}
\newcommand{\Ly}{\hat{L}_y}
\newcommand{\Lz}{\hat{L}_z}

\newcommand{\fv}{g^{(1)}_{12}}

\newcommand{\hide}[1]{}

\begin{document}

\title{Phase-diffusion dynamics in weakly coupled Bose-Einstein condensates}

\author{Erez Boukobza$^a$, Maya Chuchem$^b$, Doron Cohen$^b$, and Amichay Vardi$^a$}
\affiliation{Departments of Chemistry$^a$ and Physics$^b$, Ben-Gurion University of the Negev, P.O.B. 653, Beer-Sheva 84105, Israel}

\begin{abstract}
We study the phase-sensitivity of collisional phase-diffusion between weakly coupled Bose-Einstein  condensates, using a semiclassical picture of the two-mode Bose-Hubbard model.  When weak-coupling is allowed, zero relative phase locking is attained in the Josephson-Fock transition regime, whereas a $\pi$ relative phase is only locked in Rabi-Josephson point. Our analytic semiclassical estimates agree well with the numerical results.
\end{abstract}

\maketitle


Bose-Einstein condensates (BECs) of dilute, weakly-interacting gases are currently used as a testing ground for the quantitative study of various condensed-matter models. In addition to offering a remarkable degree of controllability, they open the way for exploring non-equilibrium {\it dynamics}, far beyond small perturbations of the ground state. Whereas  this regime is inaccessible in the equivalent condensed matter realizations, due to the high Fermi energy, highly excited states are naturally produced in BEC experiments and their dynamics can be traced with great precision.

One such example is the 'phase-diffusion' \cite{PhaseDiffusion,Vardi01,Greiner02,Schumm05,Hofferberth07,Jo07,Widera08} between two BECs prepared in a coherent state and consequently separated. Principle aspects of this process are captured by the two site Bose-Hubbard Hamiltonian (BHH).
Defining $\ei_i$, $\eidag_i$ as bosonic annihilation and creation operators for particles in condensate~${i=1,2}$, the corresponding particle number operators are $\hn_{i}=\eidag_{i}\ei_{i}$, and the BHH takes the form  \cite{BJM,Leggett01}, 
\begin{equation}
\label{Ham}
\hat{H}=\varepsilon\Lz-J\Lx+U\Lz^2~,
\end{equation}
where $\Lx=(\eidag_1 \ei_2+\eidag_2\ei_1)/2$,  $\Ly=(\eidag_1\ei_2-\eidag_2\ei_1)/(2i)$, 
and $\Lz=(\hn_1 - \hn_2)/2$, generate an $SU(2)$ Lie algebra. 
Bias potential, coupling, and collisional interaction energies are denoted as $\varepsilon$, $J$, and $U$, respectively.  We have eliminated $c$-number terms that depend on the conserved total particle number $N=\hn_1+\hn_2$.   
Below we use for representation the Fock~space basis states $|\ell,m\rangle_z$, which are the joint eigenstates of ${\hat{\bf L}}^2$ and $\Lz$, with $\ell=N/2$.


Experimental procedures allow to prepare the system in any desired $SU(2)$ coherent state ${|\theta,\varphi\rangle}=e^{-i\varphi\Lz}e^{-i\theta\Ly}|\ell,\ell\rangle_z$.  The $\theta$ rotation is realized by strong coupling $J\gg NU$,  while the phase $\varphi$ is tuned by switching the bias $\varepsilon$,  thus inducing coherent phase oscillations \cite{Schumm05,JoChoi07}.   In particular it is possible to prepare the equal-population coherent states
\begin{eqnarray}  
\label{coherento}
\left|\frac{\pi}{2},\varphi\right\rangle
&=& 
\left[
\frac{1}{\sqrt{2}}
\left(
a_1^{\dag} + \mbox{e}^{i\varphi}a_2^{\dag}
\right)^N
\right]
|\mbox{vacuum}\rangle
\\ \nonumber
~&=&2^{-\ell}\sum_{m=-l}^{\ell}\left(e^{-i\varphi}\right)^{\ell+m}
\left(
\begin{array}{c}
2l \\ \ell+m
\end{array}\right)^{1/2}|l,m\rangle_z 
\end{eqnarray}
In phase-diffusion experiments, the preparation stage is followed by a sudden separation of the condensates, 
so as to obtain two equally populated modes of a symmetric ($\varepsilon{=}0$) 
double well with a definite relative phase $\varphi$.
Due to the interaction terms, different $L_z$ Fock states 
oscillate with different frequencies, and the evolution under the Hamiltonian (\ref{Ham}) 
with $u=NU/J \gg 1$ leads to the loss of single-particle coherence, 
quantified by the multi-shot mean fringe-visibility $\fv=|\langle \Lx \rangle|/\ell$. 
For fully separated condensates the fringe visibility decays as $\fv(t)=e^{-(t/t_d)^2}$ 
with $t_d=(U\sqrt{\ell})^{-1}$ \cite{PhaseDiffusion,Greiner02,Jo07,Widera08}.
Long time revivals are obtained at $t_r=\pi/U$, as confirmed experimentally \cite{Greiner02}. 
This behavior is {\em insensitive} to the initial phase $\varphi$, 
and merely reflects the Binomial Gaussian-like distribution of the occupation, implied by Eq.~(\ref{coherento}).


In reality we still may have some inter-mode coupling~$J$, 
which is small compared with the collisional energy $NU$ \cite{Schumm05,Hofferberth07,Jo07,JoChoi07}. 
Recent experiments in 1D BECs, have explored the loss of 
single-particle coherence under such circumstances \cite{Hofferberth07}, 
resulting in the decay to a non-vanishing equilibrium value of $\fv$. We note that 
since these experiments focus on phase fluctuations associated with one dimensionality ,
their detailed description goes beyond the two-mode BHH \cite{Widera08,Gritsev06}, and 
that dephasing can also be caused by thermal noise \cite{Vardi01}. 
Other experiments have found $\varphi$-sensitivity 
in the full merging of two separated condensates \cite{JoChoi07},  
where a relative $\varphi{=}\pi$ phase led to significant 
heating losses compared to $\varphi{=}0$. The common expectation in these
experiments is that small coupling between the condensates will lead to their phase-locking
and suppress the loss of single-particle coherence.

\begin{figure}[t]
\centering
\includegraphics[width=0.50\textwidth]{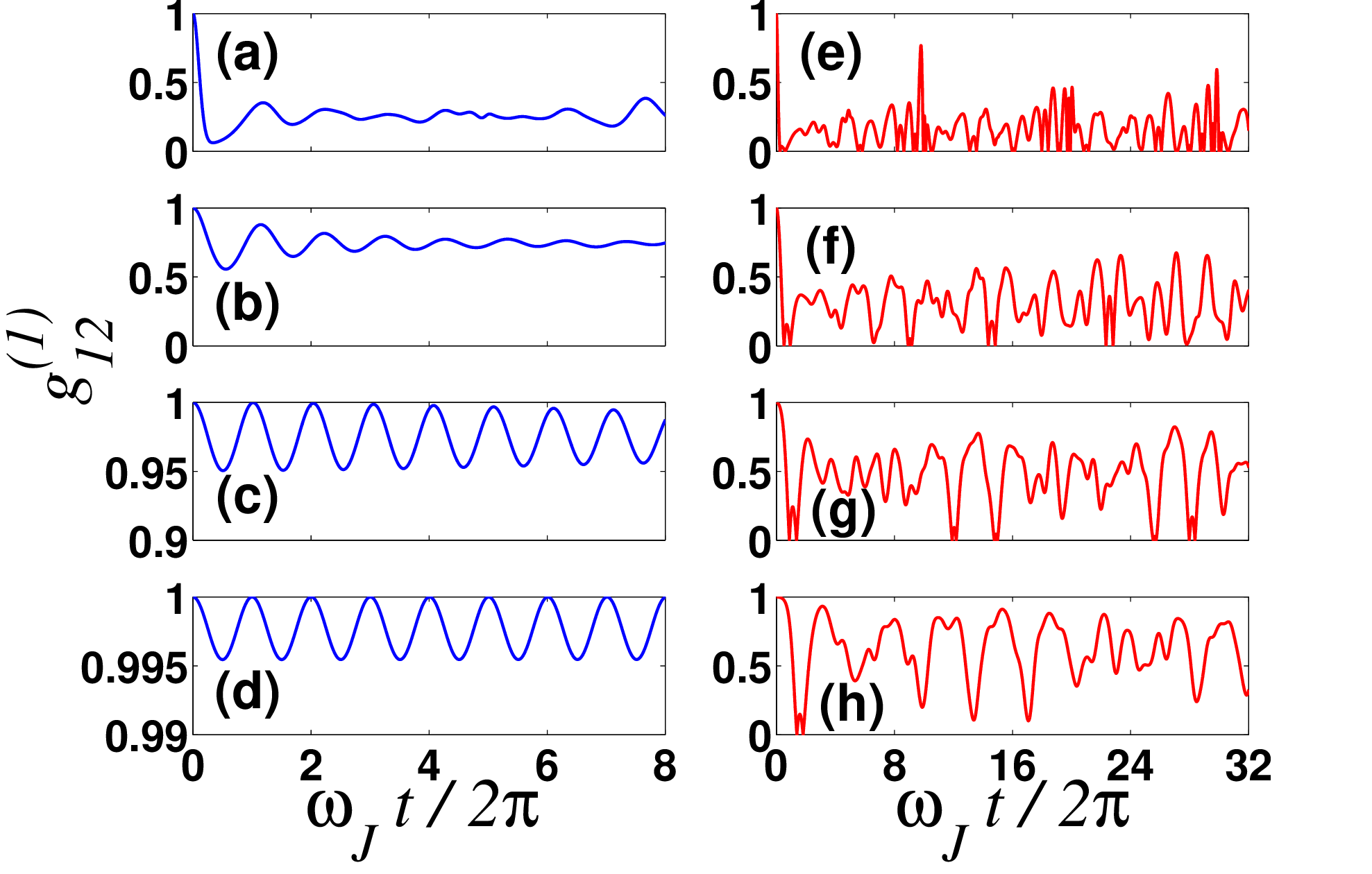}
\caption{(Color online) Evolution of fringe visibility with $N{=}1000$ particles, 
starting from the spin coherent states $|\pi/2,0\rangle$ (a-d), 
and $|\pi/2,\pi\rangle$ (e-h). 
The coupling parameter is $u=10^4$ (a,e), $10^3$, (b,f) $10^2$ (c,g), and $10$ (d,h). 
}
\label{pspd_visibility}
\end{figure}

Generally motivated by this experimental interest we here study the $\varphi$ dependence 
of the fringe visibility evolution when the two condensates remain weakly coupled 
during the hold time, assuming the prototype two-mode BHH modeling.   
In Fig.~\ref{pspd_visibility} we plot the numerically calculated $\fv$ 
for both the $|\pi/2,0\rangle$ and the $|\pi/2,\pi\rangle$ preparations.
For the $\varphi{=}0$ preparation, the expected phase-locking is obtained 
even by a very weak coupling, leading to a nonvanishing asymptotic 
value of the fringe visibility, with some small oscillations. 
However, if the system is prepared in the $\varphi{=}\pi$ state, 
phase locking is not attained if~$u$ is moderately large, and the fringe visibility 
exhibits apparently complex dynamics with a somewhat different 
characteristic timescale. The corresponding short-time dynamics was previously studied using truncated and linearized models \cite{Vardi01}. Our purpose in the present work is to provide both qualitative and quantitative longer time analysis of this phase-sensitivity of the phase-diffusion process, using a semiclassical phase-space method.

\begin{figure}[t]
\centering
\includegraphics[width=0.5\textwidth]{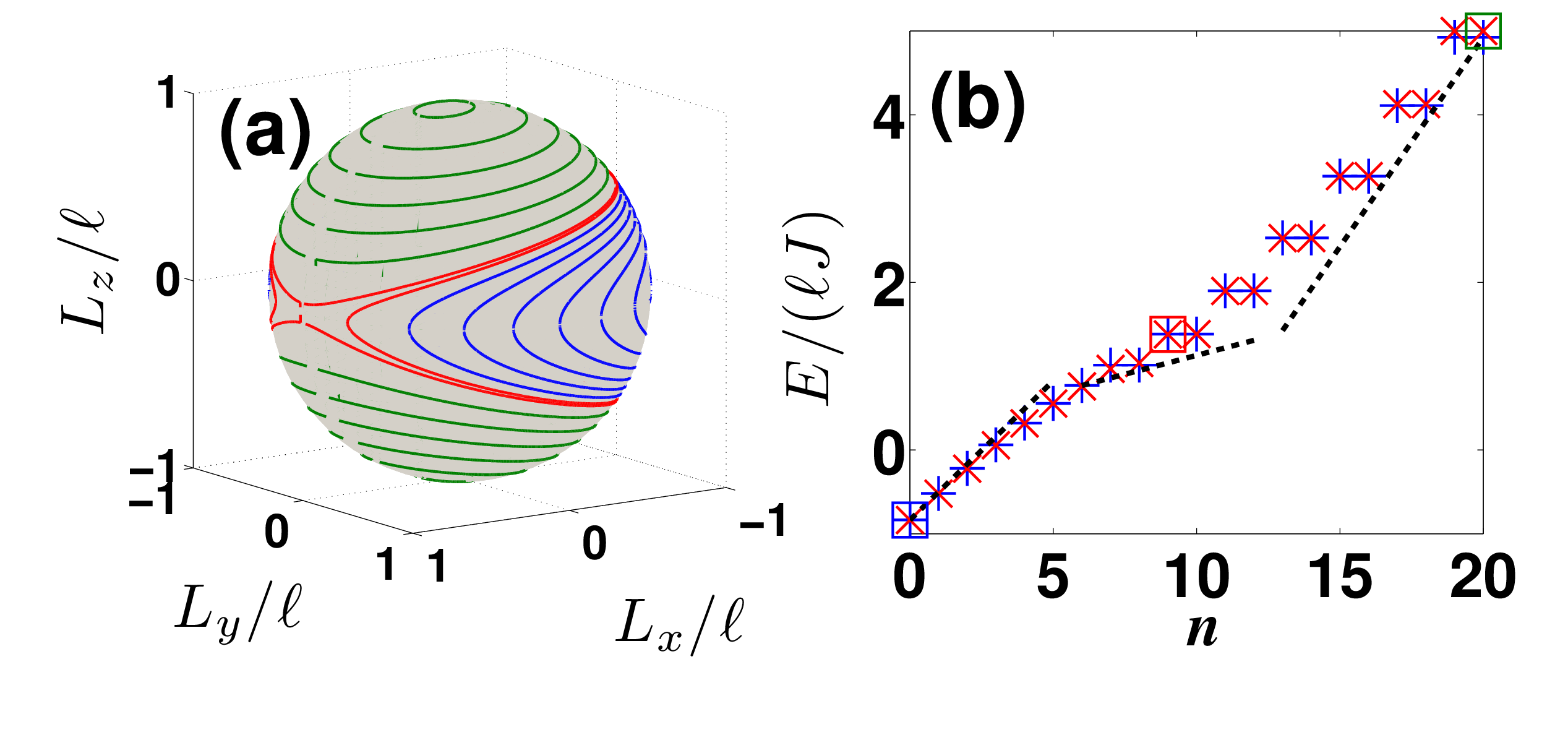}
\caption{(Color online)  Schematic drawing of the classical phase-space energy contours (a) 
and the corresponding spectrum (b), for $N={20}$ and $u=10$. WKB energies (red x) are compared with exact eigenvalues (blue +).   Dashed lines indicate slopes  $\omega_J$ for low energies, 
$\omega_x$ for near-separatrix energies, and $\omega_{+}$ for high energies. Squares mark the eigenstates plotted in Fig.~3.}
\label{pspd_phspace}
\end{figure}

{\bf Interaction Regimes -- } 
It is instructive to use a Wigner function $\rho(\theta, \phi)$  
for the representation of the quantum states \cite{agarwal}.
The Wigner function that represents a ${|\theta,\varphi\rangle}$ 
coherent state {\em resembles} a minimal Gaussian centered 
at the corresponding angle on the BHH's spherical phase space.  
Using semiclassical reasoning one can argue that the Wigner functions 
of the BHH eigenstates (with ${\varepsilon=0}$) 
are concentrated along the contour lines 
of the Gross-Pitaevskii classical energy functional,
\begin{eqnarray}  \label{e3}
E(\theta,\varphi) =   
\frac{NJ}{2}\left[\frac{1}{2} u (\cos\theta)^2 - \sin\theta\cos\varphi\right]
\end{eqnarray}
Furthermore, within the framework of the WKB approximation \cite{item10,item11} 
the quantization of the energy is implied by the condition ${A(E_n)=(4\pi/N)(n+1/2)}$ 
where $A(E)$ is the phase-space area enclosed by a fixed energy~$E$ contour, 
and $4\pi/N$ is the Planck cell. [Below we measure $A(E)$ in Planck cell units]. 
The phase space picture implies that one has to distinguish 
between three regimes according to the value of the dimensionless 
interaction parameter ${u\equiv NU/J}$ \cite{BJM,Leggett01}: 
{\bf (a)}~The linear, weak-interaction {\it Rabi} regime ${u<1}$; 
{\bf (b)}~the intermediate {\it Josephson} regime ${1<u<N^2}$; 
and {\bf (c)}~the extreme strong-interaction {\it Fock} regime ${u>N^2}$.
In the Josephson regime the spherical phase-space is split
by a figure-eight separatrix trajectory (see Fig~\ref{pspd_phspace}), 
to a 'sea' of Rabi-like (blue) trajectories 
and two interaction-dominated nonlinear 'islands'  (green).  
In the Fock regime the phase-space area of the sea becomes 
smaller than Planck cell, and therefore effectively disappears.

{\bf Spectrum -- }
The energy landscape for $u\gg1$, as implied by Eq.(\ref{e3}), 
consists of sea levels that extend from the bottom energy ${E_{-}=-\ell J}$, 
and of island levels that occur in almost degenerate pairs and extend  
up to the upper energy $E_{+} \approx \ell^2U$. The border between 
the sea and the islands is the separatrix energy ${E_{\rm_x}=\ell J}$. 
The oscillation frequency $\omega(E)=[A'(E)]^{-1}$ 
around the minimum-energy (stable) fixed point ${(\pi/2,0)}$ is the 
plasma / Josephson frequency  
\begin{eqnarray}  
\omega_J \ \equiv \ \omega(E_{-})  \ = \ \sqrt{(J+NU)J} \ \approx \ \sqrt{NUJ}
\end{eqnarray}
If we had $u<1$ then the opposite phase-space fixed point ${(\pi/2,\pi)}$  would be also 
stable with $\omega_{+}=\sqrt{(J-NU)J}$. But for ${u>1}$ it bifurcates, 
and replaced by an unstable fixed point at the separatrix energy, 
accompanied by the two twin stable fixed points $(\arcsin(1/u),\pi)$,   
located within the islands. For $u\gg1$ these stationary points approach    
the poles ($\arcsin(1/u)=0,\pi$) and the oscillation frequency becomes $\omega_{+} \approx NU$. 
The significance of the various frequencies is implied by the WKB quantization (Fig.~2). 
At low-energy we have a non-degenerate set of Josephson levels
with spacing  $\omega_J$. Due to the non-linearity the spacing becomes 
smaller as one approaches the separatrix energy (see next paragraph), 
while the high-energy levels (${E_m=Um^2}$) are doubly-degenerate 
with spacing $2Um$ that approaches $\omega_{+}$ as $m\rightarrow\ell$.

{\bf Nonlinearity -- }
The motion in the vicinity of the separatrix is highly non-linear.
This is implied by the non-linear variation of the phase space area near the separatrix: 
\begin{eqnarray}  
|A(E)-A(E_{\rm x})|= \left|\frac{E{-}E_{\rm x}}{\omega_J}\right|\log\left|\frac{NJ}{E{-}E_{\rm x}}\right|
\end{eqnarray}
We stress that Planck cell units are used for $A(E)$. 
On the basis of this expression, the WKB quantization condition implies 
that the level spacing at the separatrix is {\em finite} 
and given by the expression   
\begin{eqnarray}  \label{e4}
\omega_{\rm x} \ \ = \ \ \left[\frac{1}{2}\log\left(\frac{N^2}{u}\right)\right]^{-1} \omega_J
\end{eqnarray}
It is important to notice that in the strict classical limit 
this frequency becomes zero \cite{item14}, 
while for finite $N$ it differs from the Josephson frequency only by a logarithmic factor. 
The non-linearity can be characterized by a parameter $\alpha$ 
that reflects the $E$ dependence of the oscillation
frequency $\omega(E)$, and hence upon WKB quantization 
it reflects the dependence of the level 
spacing $E_{n{+}1}-E_{n} \approx \omega(E_n)$ 
on the running index $n$. The implied definition 
of this parameter and its value in the vicinity 
of the separatrix are: 
\begin{eqnarray}  \label{e5}
\alpha(E)  \equiv  \frac{1}{\omega}\frac{d\omega}{dn} 
= \omega(E)^2 A''(E) 
\sim \left[\log\left(\frac{N^2}{u}\right)\right]^{-1} 
\end{eqnarray}
Note that this parameter is meaningful only 
deep in the Josephson regime, where it is less than unity. 
As shown below, the parameter $\alpha$ controls 
the amplitude of the fluctuations in the case of a $\varphi{=}\pi$ preparation. 
Strangely enough, in the classical limit ($N\rightarrow\infty$ with fixed $u$) 
the non-linear effect becomes more pronounced 
as far as Eq.(\ref{e4}) is concerned, 
but weaker with regard to Eq.(\ref{e5}).

\begin{figure}[t]
\centering
\includegraphics[clip,width=0.50\textwidth]{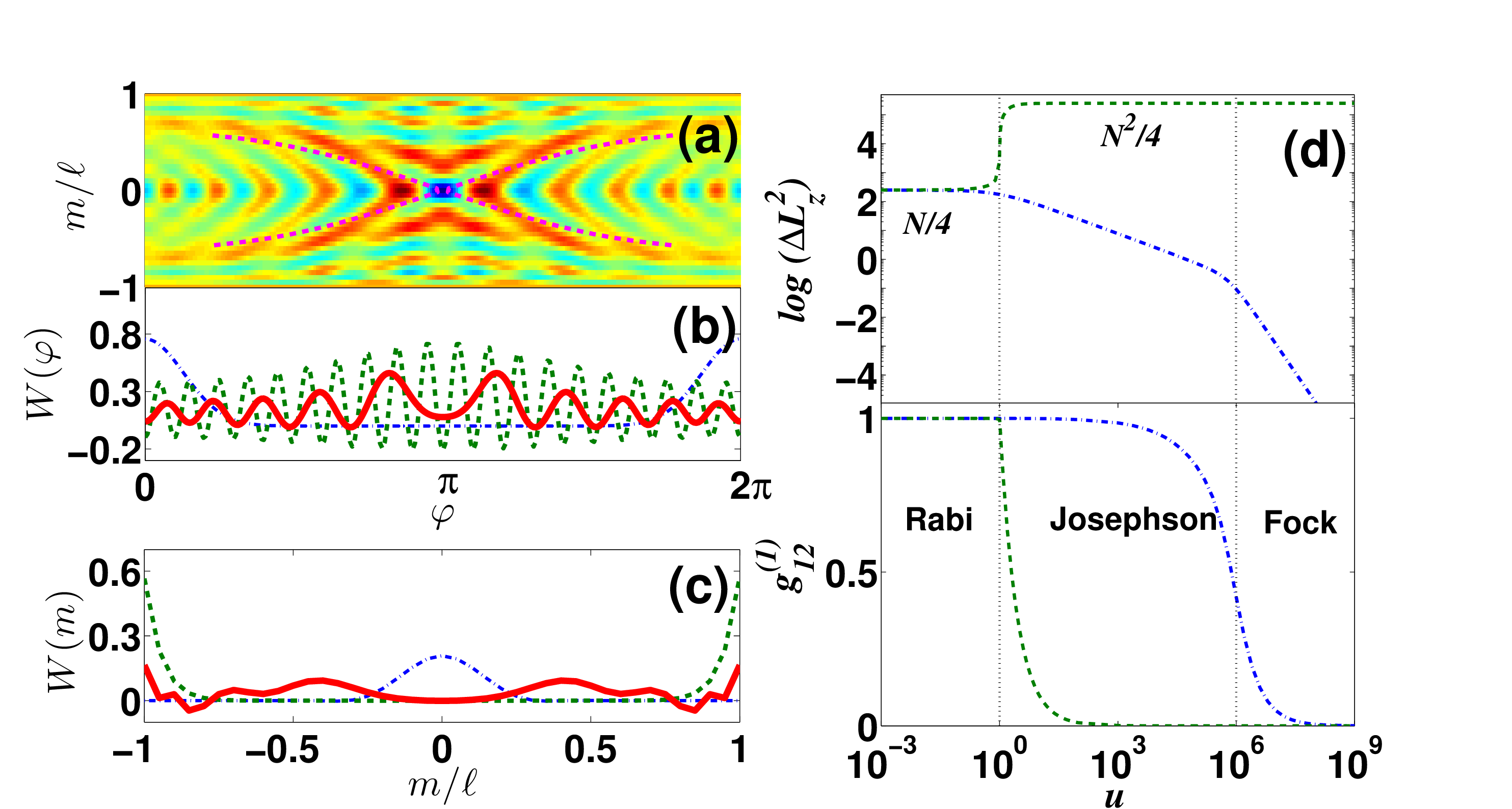}
\caption{(Color online) Wigner function of the separatrix state (a) and its $n$ and $\varphi$ projections for the eigenstates specified in Fig.~2 (b,c). Solid red lines depict the near separatrix eigenstate, while dash-dotted blue and dashed green lines correspond to the ground and most excited ('cat') states, respectively. Also shown in (d) are the relative-number variance and the fringe-visibility for ground state (blue dash-dotted line) and most excited state (dashed green line) with $N{=}1000$.}
\label{pspd_grstate}
\end{figure}

{\bf Eigenstates -- } 
(Fig.~\ref{pspd_grstate}) 
In the Rabi regime, the eigenstates of Hamiltonian approach the $\Lx$ eigenstates $|\ell,m'\rangle_x$, with eigenvalues $-Jm'$ proportional to the relative number difference between the even and odd superpositions of the modes. In particular, the lowest eigenstate $|\ell,\ell\rangle_x$ and the highest eigenstate $|\ell,-\ell\rangle_x$, are the coherent states $|\pi/2,0\rangle$ and $|\pi/2,\pi\rangle$, respectively, with binomial $m$ distributions in the $|\ell,m\rangle_z$ basis, approaching the normal (Gaussian/Poissonian) distribution for large $\ell$. 
As $u$ increases beyond unity, a transition is made into the Josephson regime. The relative-number variance $\Delta L_z$ of the ground state decreases continuously in this regime from its coherent $\sqrt{N}/2$ value. However, the coherence $\fv$ of the ground state remains close to unity and the relative-phase is well-defined throughout the Josephson regime, justifying the use of mean-field theory to depict the Josephson dynamics of ground state perturbations. The coherence of the ground state is only lost in the Fock regime, where the true ground state approaches the relative site-number state $|\ell,0\rangle_z$.

{\bf $\varphi{=}0$ preparation -- } 
The above discussion implies that in the case of a coherent preparation $|\pi/2,0\rangle$, any weak coupling beyond $J>U/N$ would suffice to {\em lock} the relative phase and to prevent the loss of single-particle coherence \cite{Hofferberth07}. Using a phase space language this means that the  $\varphi{=}0$ preparation is located at the minimum of the sea region, around a stable fixed point ${(\pi/2,0)}$ , and resembles the ground state of the Hamiltonian. The expected small-oscillation frequency for the evolution of this coherent state is $\omega_J$, and hence that of the visibility is $2\omega_J$ (see Fig.~4).

{\bf $\varphi{=}\pi$ preparation -- } 
For the coherent preparation $|\pi/2,\pi\rangle$ the picture is quite different. 
In the Rabi regime, this state coincides with the $|\ell,-\ell\rangle_x$ most excited eigenstate, 
and hence it is possible to have phase locking around the stable fixed point ${(\pi/2,\pi)}$, 
where the characteristic frequency is $\omega_{+}$. 
However, as soon as $u>1$, this fixed point bifurcates, and the Wigner function 
of the eigenstates has to equally populate the two twin islands in phase space, 
hence forming a cat state \cite{Cat} with bimodal $m$~distribution.   
This crossover is reflected (see Fig.~\ref{pspd_grstate})  
by the sharp increase in the relative-number variance 
to the cat-state value $\Delta L_z=N/2$, 
and by the equally sharp decrease of its single-particle coherence.
The evolution of the coherent preparation $|\pi/2,\pi\rangle$ in the vicinity 
of the unstable point is further analyzed below.

\begin{figure}[t]
\centering
\includegraphics[clip,width=0.5\textwidth]{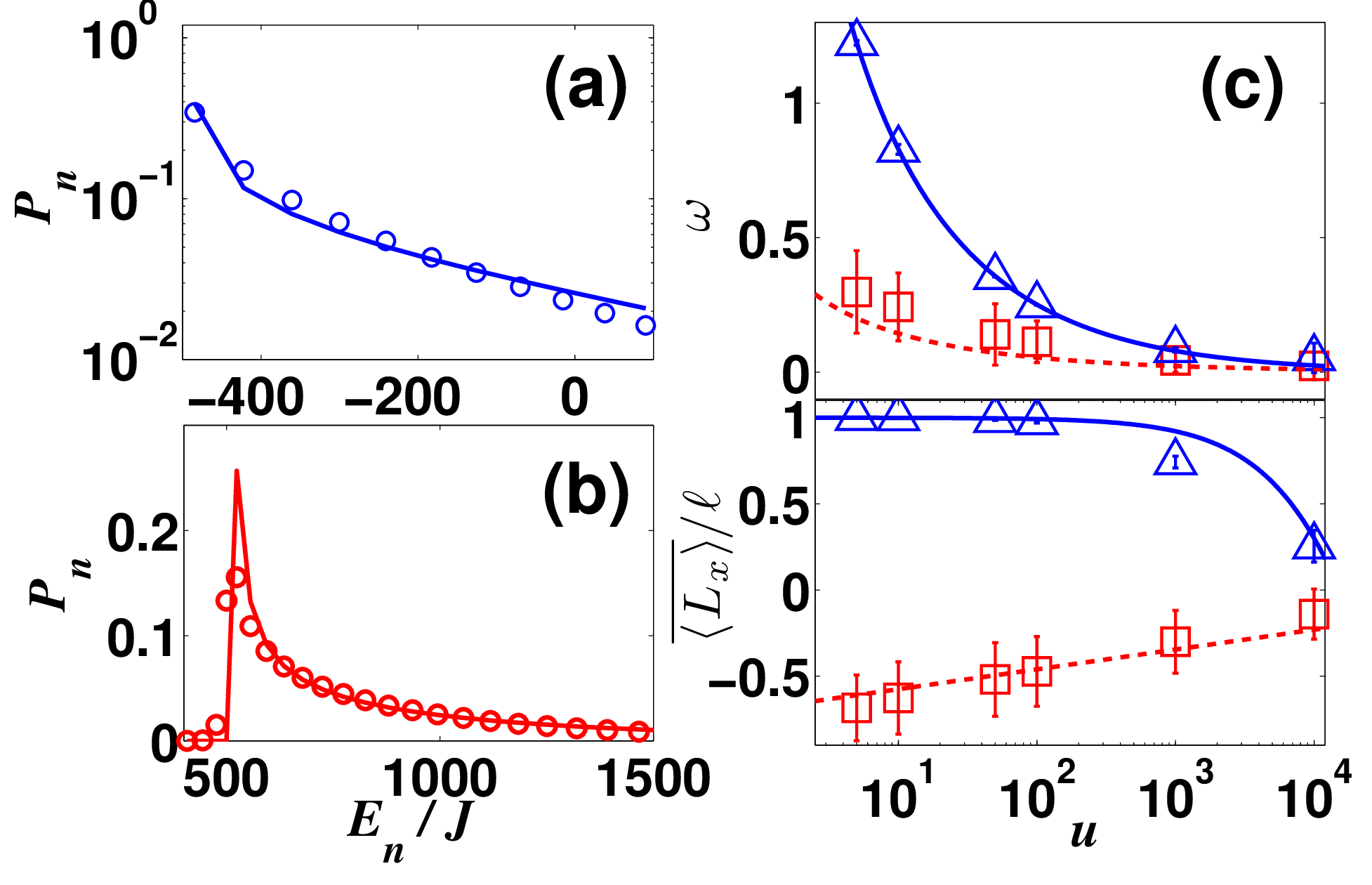}
\caption{(Color online) 
Probability distribution $\mbox{P}(E_n)$ for the preparations $|\pi/2,0\rangle$ (a)
and $|\pi/2,\pi\rangle$ (b), with $N{=}1000$ particles, and $u=10^3$.  Numerical 
expansions ($\circ$) are compared with the semiclassical estimates of
Eq.(\ref{LDOSz}) and Eq.(\ref{LDOSpi}) (solid lines). In (c), the mean frequency and the 
long time average of the $\langle L_x \rangle$ oscillations   
in Fig.~1(b,f) ($\triangle$ for $|\pi/2,0\rangle$, $\Box$ for $|\pi/2,\pi\rangle$) are compared to
the semiclassical analysis of the dynamics in phase space (solid blue and dotted red lines, respectively).
Error bars reflect the associated dispersion. 
}
\label{pspd_ldos}
\end{figure}

{\bf Dynamics -- } 
The phase-sensitivity of phase-diffusion originates from the instability of the hyperbolic fixed point $(-1,0,0)$ of the classical phase space, as opposed to the stability of the elliptic fixed point $(1,0,0)$  \cite{Vardi01}. Here we would like to go beyond this heuristic observation and understand quantitatively the long time recurrences in Fig.~\ref{pspd_visibility}. 
The coherent state $|\pi/2,0\rangle$ is a superposition of a few low-energy states. Since the level-spacing is fixed on the Josephson frequency $\omega_J$, the fringe visibility carries out clean harmonic oscillations. 
By contrast, the state $|\pi/2,\pi\rangle$ consists of levels around the separatrix energy, which are not equally spaced, 
resulting in the quasi-periodic oscillations of Fig.~\ref{pspd_visibility}(e)-(h). 
The observed timescale of these oscillations (Fig.~4) agrees well with $\omega_{\rm x}$.
For the purpose of quantitative analysis  
we have to expand the initial state $\psi$ in the $E_n$ basis. 
A reasonable estimate for the envelope function 
$P(E_n) = |\langle E_n | \psi\rangle|^2 = \mbox{trace}(\rho^{(n)} \rho^{(\psi)})$ 
can be obtained using a semiclassical approximation. 
The Wigner function of the $n$th eigenstate is approximated by a microcanonical distribution   
$\rho^{(n)}(\varphi,\theta) =  \omega(E_n) \delta(E(\theta,\varphi)-E_n)$ and the coherent 
state is approximated by a minimal Gaussian. 
With appropriate approximations for $u\gg1$ the phase space integration gives  
for the $\varphi{=}0$ preparation
\begin{equation}
\mbox{P}({\tilde E}) 
= 2\mbox{\bf I}_0\left[\left(2{-}\frac{1}{2u}\right){\tilde E}\right]
e^{-\left(2{+}1/2u\right){\tilde E}}
\label{LDOSz}
\end{equation}
with $\tilde E\equiv (E-E_-)/J$, and for the $\varphi{=}\pi$ preparation
\begin{equation}
\mbox{P}({\tilde E}) = 
\frac{1}{\pi}\left(\frac{\omega(E)}{\omega_J}\right)
\mbox{\bf K}_0\left[\left(2{+}\frac{1}{2u}\right)|{\tilde E}| \right]
e^{\left(2{-}1/2u\right) {\tilde E}}
\label{LDOSpi}
\end{equation}
with ${\tilde E}\equiv(E-E_{\rm x})/J$.  In the above, ${\rm\bf I_0}$ and ${\rm\bf K_0}$ are the modified Bessel functions.  As shown in Fig.~\ref{pspd_ldos}(a) and Fig.~\ref{pspd_ldos}(b) this analytical estimate agrees remarkably well with the exact numerical diagonalization, with the exception of the separatrix state whose phase-space distribution (Fig. \ref{pspd_grstate}(a)) somewhat deviates from the microcanonical ansatz. Due to the even parity of both preparations ($|\pi/2,0\rangle$ is always even and $|\pi/2,\pi\rangle$ has $(-1)^N$ parity), the occupation of odd $n$ states vanishes and the occupation of the even $n$ states is twice the semiclassical estimate.  Using these 
expressions we can estimate the participation ratio of the preparation, 
which comes out $M \sim NU/\omega_J\approx\sqrt{u}$.

{\bf Fluctuations -- } 
Regarding $\psi$ as a superposition of $M$ energy states 
we can estimate the fluctuations using the following reasoning:   
In the linear case, if the energy levels are equally spaced,  
there is only one ${\mathcal{M}=1}$ basic frequency  
and the motion is strictly periodic.
If the energies were quasi-random, then 
there are $\mathcal{M}=M{-}1$ independent frequencies, 
and the motion is quasiperiodic with 
relative variance $\sim 1/\mathcal{M}$. 
But what we have for the $\pi$ preparation are 
energy levels that are characterized 
by a nonlinearity parameter $\alpha$.   
Then, if the nonlinearity parameter is not very small,  
the effective number of independent frequencies 
is $\mathcal{M} = \sqrt{\alpha} M$, which interpolates 
between the linear and the quasi-random estimates. 
As shown in Figs. 4(c) and 4(d), this analysis is consistent with 
our numerical simulation.

{\bf Summary -- }  
The vast majority of recent BEC interference experiments related to phase-diffusion, are carried out in the Josephson regime. Within it, the dynamics of single-particle coherence is expected to strongly depend on the initial relative-phase between the partially separated condensates. Zero relative phase leads to phase-locking even for $u$ as large as $N^2$, whereas a $\pi$ relative phase preparation results in the loss of fringe-visibility all the way down to $u\sim 1$. Using a semiclassical phase-space picture we related this behavior to the classical phase-space structure and the ensuing WKB spectrum, obtaining analytical expressions for the envelope functions of these two coherent preparations in the eigenstate basis, for $u$ in the Josephson regime. These expansions agree well with numerical calculations and enable the accurate prediction of the amplitude of fringe-visibility oscillations.


This work was supported by the Israel Science Foundation (Grant 582/07), 
and by a grant from the USA-Israel Binational Science Foundation (BSF).

\vspace*{-4mm}



\end{document}